\documentclass[twocolumn,showpacs,prb,floatfix]{revtex4}
\usepackage{epsfig}

\newcommand{\be}{\begin{equation}}
\newcommand{\ee}{\end{equation}}
\newcommand{\beqn}{\begin{eqnarray}}
\newcommand{\eeqn}{\end{eqnarray}}

\begin{document}
\noindent


\title{Dynamic scaling in spin glasses}


\author{C. Pappas}
 \email{pappas@hmi.de}
\affiliation{HMI Berlin, Glienickerstr. 100, 14109 Berlin, Germany}

\author{F. Mezei}
 \email{mezei@hmi.de}
\affiliation{HMI Berlin, Glienickerstr. 100, 14109 Berlin, Germany}
\affiliation{Los Alamos National Laboratory, Los Alamos USA}

\author{G. Ehlers}
 \email{ehlers@ill.de}
\affiliation{ ILL, 38042 Grenoble Cedex 9, France}

\author{ P. Manuel}
\email{P.Manuel@rl.ac.uk}
\affiliation{Physics \& Astronomy Department, University of Leeds LS2 9JT UK
\footnote {present address : ISIS, Rutherford Appleton Laboratory, Chilton, Didcot OX11 OQX, UK}}

\author{I.A. Campbell}
\email{campbell@ldv.univ-montp2.fr}
\affiliation{Laboratoire des Verres, University Montpellier II, place
Eug\`{e}ne Bataillon 34095 Montpellier Cedex 5, France}

\begin{abstract}

We present new Neutron Spin Echo (NSE) results and a revisited analysis of historical data on spin glasses, which reveal a pure power-law time decay of the spin autocorrelation function $s(Q,t) = S(Q,t)/S(Q)$ at the glass temperature $T_g$, each power law exponent being in excellent agreement with that calculated from dynamic and static critical exponents deduced from macroscopic susceptibility measurements made on a quite different time scale. It is the first time that this scaling relation involving exponents of different physical quantities determined by completely independent experimental methods is stringently verified experimentally in a spin glass. As spin  glasses are a subgroup of the vast family of glassy systems also comprising structural glasses and other non-crystalline systems the observed strict critical scaling behaviour is important. Above the phase transition the strikingly non-exponential relaxation, best fitted by the Ogielski (power-law times stretched exponential) function, appears as an intrinsic, homogeneous feature of spin glasses.
\end{abstract}
\maketitle

%

The glass transition, characterised by a dramatic slowing down of the dynamics without any noticeable change in the spatial order, is a generic phenomenon, seen in systems as different as disordered magnets, polymers and biological substances. In spite of its universality and of intense experimental and theoretical efforts it is still controversial whether the glass transition in structural glasses is a gradual freezing or a phase transition. The difficulty to identify the nature of the glass transition is due to the absence of an observable order parameter analagous to magnetization in the low temperature phase, usually a key quantity in the study of phase transitions.  This is due to the absence of any static spacial fingerprint; instead, the  order parameter appears in the dynamics \cite{parisi, goetze}. In fact, the "snapshot" structure factor $S(Q) = S(Q, t=0)$, which reflects the short and medium range static correlations, shows no essential change when passing from the high temperature liquid (or paramagnet) to the low temperature frozen glass phase. In this situation the observation of dynamic scaling relations, which are the direct consequence of the homogeneity hypothesis in the vicinity of a critical instability \cite{widom} can reveal the crucial signature of a true phase transition.

In spin glasses, which are the simplest realisations of glassy systems from the experimental as well as from the theoretical point of view, a phase transition is well established\cite{fischer,young}. Low frequency dynamic susceptibility as well as the non linear part of the static susceptibility follow the scaling relations and the analysis leads to accurate determination of the static critical exponents $\gamma$, $\beta$, $\delta$ and of the dynamic exponent $z$ \cite{levy, nadine, paulsen, mydosh}. The verification of scaling relations between certain exponents, determined by completely independent experimental methods, however, was up to now impossible due to the absence of any obvious critical behaviour on other physical quantities, like the specific heat. Here we show that  scaling relations can also be verified experimentally in spin glasses. New Neutron Spin Echo (NSE) results and also a revisited analysis of historical data\cite{feri-murani, feri, eusrs} show a pure power-law time decay of the spin autocorrelation function $s(Q,t) = S(Q,t)/S(Q, t=0)$ at the glass temperature $T_g$, with a power law decay exponent which is in excellent agreement with that calculated from the dynamic and static critical exponents deduced from zero and low frequency susceptibility measurements. The interplay between neutron scattering, macroscopic magnetic, hyperfine field measurements, and simulations has always been decisive in understanding spin glasses and these results constitute the strongest experimental evidence yet for a true phase transition with a non-conventional order parameter; they also imply that the prominently non-exponential relaxation is an intrinsic, homogeneous feature of spin glasses.

       %
       %
\begin{figure}[h]
\resizebox{0.5\textwidth}{!}{%
\includegraphics[scale=0.25]{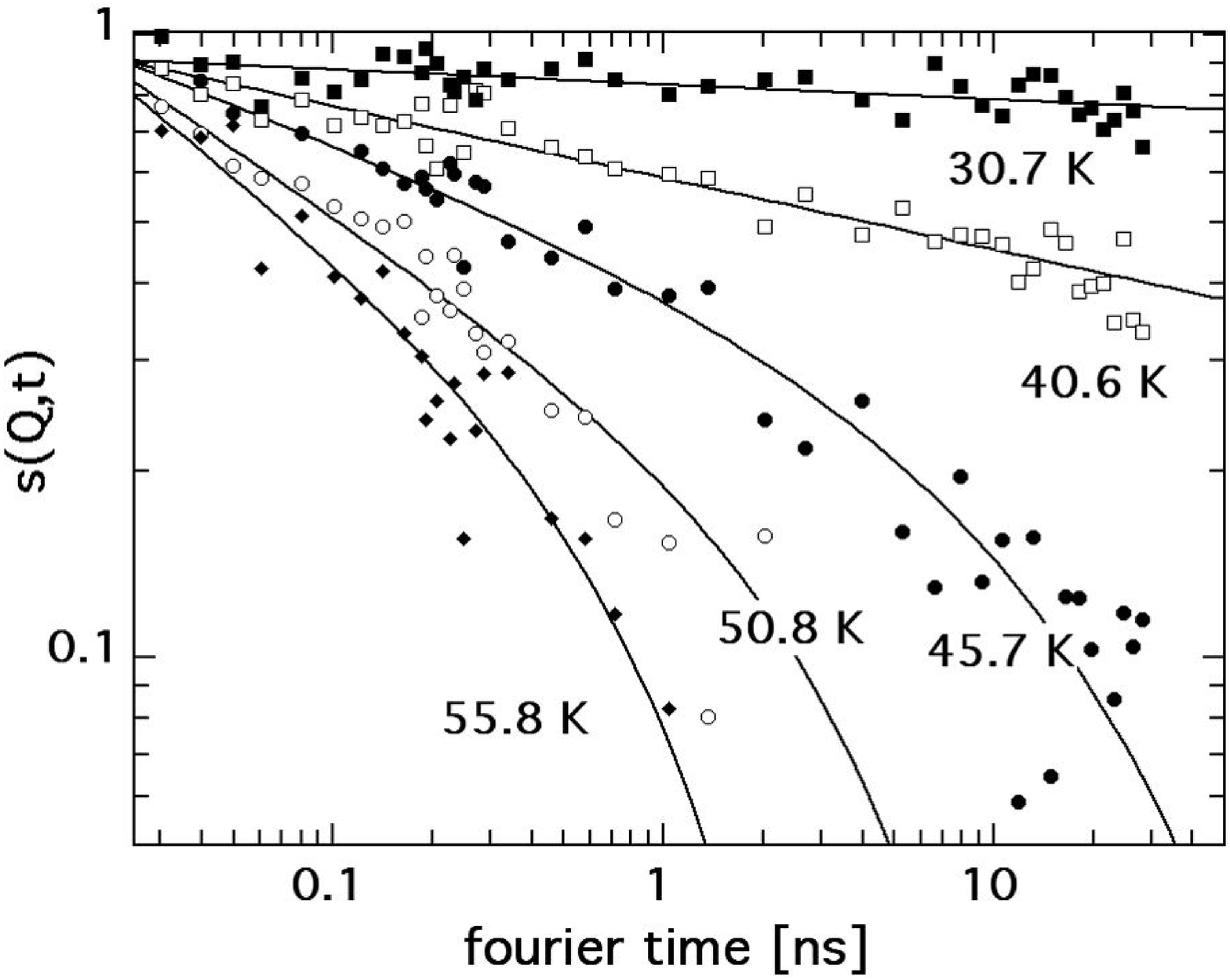}
}
\caption{Temperature dependence of the normalised intermediate scattering function s(Q,t) of Au$_{0.86}$Fe$_{0.14}$. The spectra were collected at Q = 0.4{\nobreakspace}nm$^{-1} $ with the Neutron Spin Echo spectrometer IN15 (ILL) for T=30.7{\nobreakspace}K (closed squares), 40.6{\nobreakspace}K (${\sim}$T$_{g}$, open squares), 45.7{\nobreakspace}K (closed circles),
50.8{\nobreakspace}K (open circles) and 55.8{\nobreakspace}K (closed rhombs) respectively. The continuous lines are the best fits to the data of a simple power law decay below T$_{g}$ (${\sim}$41{\nobreakspace}K) and of the Ogielski function above T$_{g}$ (see text)}
\label{all in15 data}       
\end{figure}
                               %

Most of the information about glass transitions comes from the high temperature unfrozen phase, where thermodynamic equilibrium is easily reached without any long-time drifts and aging phenomena. In spin glasses, where the magnetization is always zero, the fundamental parameter is the mean spin autocorrelation function
$q(t-t^{'})=<\bf{S}_i(t).\bf{S}_i(t^{'})>$ where $\bf{S}_{i}$ is the spin at a site $i$ and the average runs over all sites and configurations of the sample. Critical behaviour in the paramagnetic phase is seen in the non-linear susceptibility (or "spin glass susceptibility"). \cite{levy, nadine, paulsen, mydosh} Below $T_g$ the Edwards Anderson order parameter
   q(t${\rightarrow}{\infty})= \texttt{<}\bf{S}_{i}(t=0)\bf{S}_{i}(t{\rightarrow}{\infty})\texttt{>}$,
becomes non-zero.\cite{edwards-anderson, parisi} Neutron Spin Echo spectroscopy measures the scattering function $S(Q,t)$ and after normalization by $S(Q,0)$ delivers a direct determination of the autocorrelation function $q(t)=s(Q{\rightarrow}0,t)$. NSE   covers a time domain ranging from 10$^{-12}${\nobreakspace}s to some 10$^{-8}${\nobreakspace}s, i.e. from characteristic microscopic times up to times, which already belong to the "long" time relaxation domain. The first NSE experiment ever performed on a glassy system was made on the reference spin glass CuMn 5\% in 1979 \cite{feri-murani} and the results strongly influenced subsequent thinking on (spin)glass dynamics.  For the first time is was shown  that non-conventional dynamics is not limited to the spin glass phase but also extends into the paramagnetic phase well above T$_{g}$. Non-exponential and Q-independent relaxation occurs in a large temperature range up to 2-3 T$_{g}$, which can arguably be identified with the Griffiths phase \cite{griffiths}.  For about T$>$1.2 T$_{g}$ the relaxation can be described by a broad distribution of Arrhenius activation energies. Closer to T$_{g}$, however, a more dramatic slowing down sets in, which can be interpreted as the footprint of a phase transition with a critical region of usual extent.  Here we report on the first detailed analysis of $s(Q,t)$ around T$_{g}$ in spin glasses, based on enhanced quality data obtained by using new generation NSE spectrometers.

%
%
%
%
\begin{figure}[t]
\resizebox{0.5\textwidth}{!}{%
\includegraphics[scale=0.33]{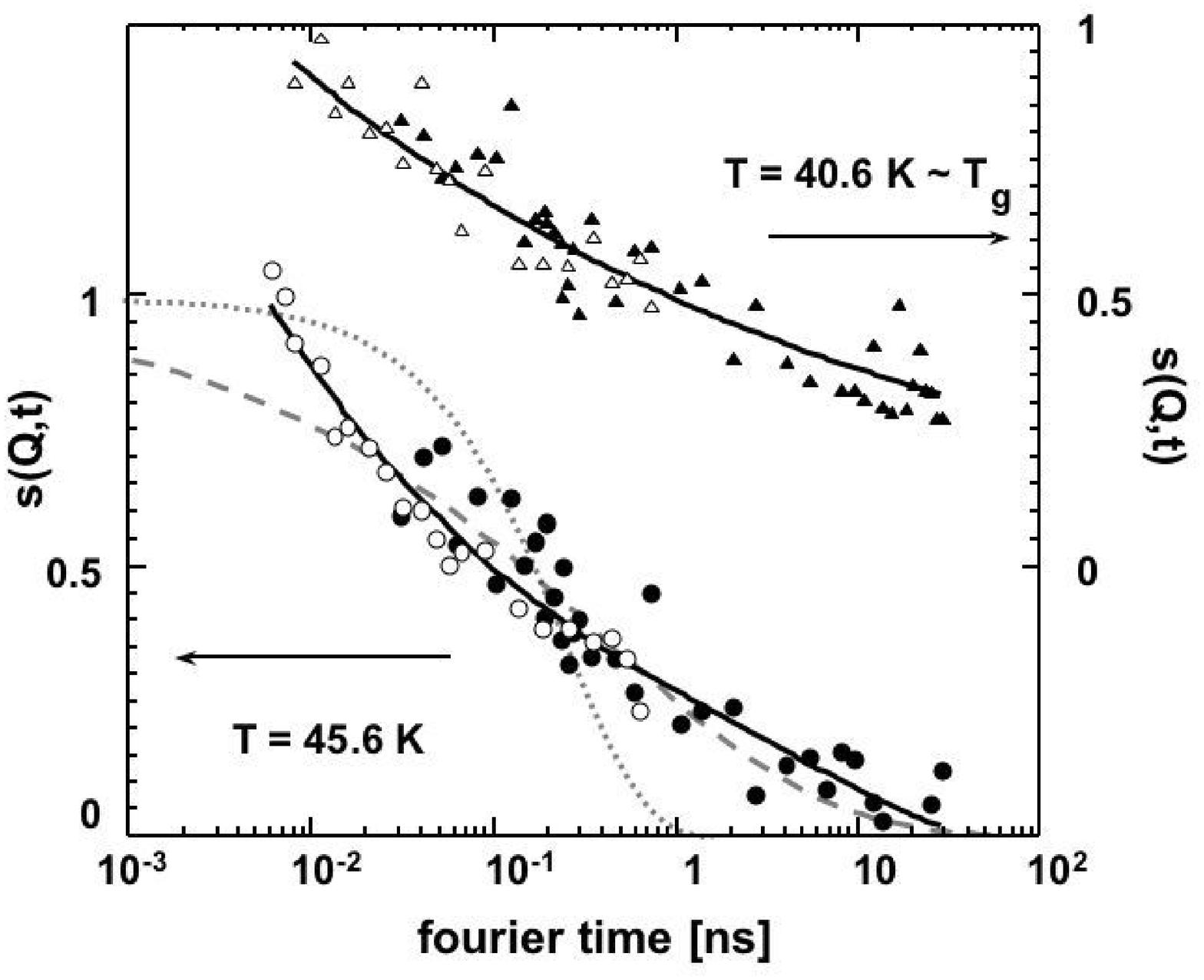}
}
\caption{NSE spectra of Au$_{0.86}$Fe$_{0.14}$ collected at  Q=$0.08 \AA^{-1}$ at the ILL 
spectrometer IN15 (full symbols) and at the BENSC spectrometer SPAN 
(open symbols) plotted in a lin-log scale. The circles are measured at 45.6 K and the triangles at 40.6 K respectively. 
For the sake of clarity the data sets corresponding to each of the  temperatures have been shifted with respect to each other   in the vertical scale.
The data at 40.6 K are fitted to a simple power law (continuous line). 
The continuous line through the 45.6 K data represents the fit of an Ogielski  function (see text). 
The dashed and dotted curves correspond to a stretched exponential and a simple exponential decay respectively. }

\label{span-in15}       
\end{figure}
%

For an accurate determination of the NSE spectra we chose Au$_{0.86}$Fe$_{0.14}$. AuFe is a classical metallic Heisenberg spin glass with significant local anisotropy \cite{dorothee} and with strong ferromagnetic correlations which amplify the magnetic scattering in the forward direction so improving the ratio between the magnetic signal and all non-magnetic (structural) contributions, i.e. the signal to noise ratio. The sample was a polycrystalline disc 0.5{\nobreakspace}mm thick with a diameter of 37{\nobreakspace}mm prepared by arc melting of the constituents. It was subsequently cold worked, homogenised at 900$^{o}$C, annealed at 550$^{o}$C and then quenched and kept in liquid Nitrogen. \cite{aufe_prep} Given the vicinity to the percolation threshold x$_{c}$ (${\sim}$15.5\% at. Fe), above which ferromagnetism sets in,\cite{coles}  the annealing and quenching procedure was repeated before every series of measurements. The NSE data were collected at the high resolution spectrometer IN15 of ILL \cite{in15} at an incoming wavelength of 0.8{\nobreakspace}nm for Q = 0.4 and 0.8{\nobreakspace}nm$^{-1} $  respectively. These results were supplemented by measurements at the wide angle NSE spectrometer SPAN of BENSC \cite{span} at an incoming wavelength of 0.45{\nobreakspace}nm for 0.6{\nobreakspace}nm$^{-1} {\leq}$Q ${\leq}$ 2.6{\nobreakspace}nm$^{-1}$. We used the paramagnetic NSE set-up, which directly delivers the magnetic part of the NSE signal and for this reason no background correction was required. All NSE spectra were normalized against the resolution function of the spectrometers, determined with the sample well below T$_{g}$, at 2{\nobreakspace}K, where the spin dynamics is completely frozen. A small part of the sample was taken out for dc susceptibility measurements with a commercial SQUID magnetometer at the HMI and for ac susceptibility measurements, which were made from 10 Hz up to 10 KHz with the MAGLAB setup at the Physics and Astronomy Department, University of Leeds. The spin glass temperature of $T_g=41.0\pm 0.3$ K was determined from the maximum of the static susceptibility.

%
%
\begin{figure}[t]
\resizebox{0.48\textwidth}{!}{%
\includegraphics[scale=0.29]{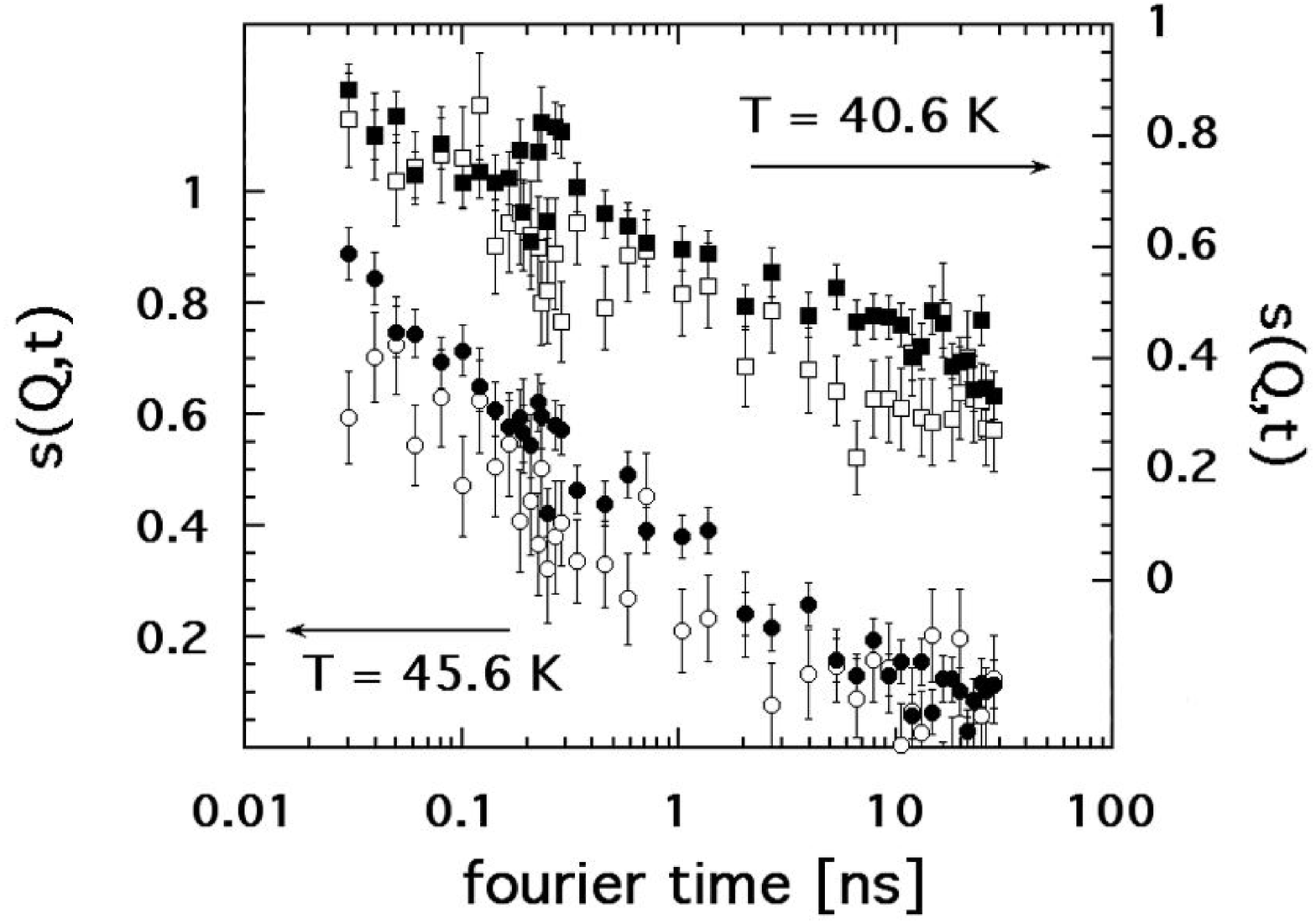}
}
\caption{s(Q,t)  of Au$_{0.86}$Fe$_{0.14}$ measured at 0.04 
\AA$^{-1}$ (close symbols)
and 0.08 \AA$^{-1}$ (open symbols) for 45.6 and 40.6 K respectively. 
For the sake of clarity the 
data sets corresponding to each of the  temperatures have been shifted with 
respect to each other   in the vertical scale. }
\label{q-dependence}       
\end{figure}
%

The normalised intermediate scattering function s(Q,t) of Au$_{0.86}$Fe$_{0.14}$ at Q = 0.4{\nobreakspace}nm$^{-1} $ is shown in Fig. 1 plotted in a log-log scale. The spectra span a dynamic range of three orders of magnitude and  by combining spectra collected at two wavelengths on IN15 and SPAN the time domain of the observation 
is extended up to almost 4 decades (Fig.2). The time dependence of the experimental $s(Q,t)$ is impressively similar to that of the numerical $q(t)$ found in large scale Ising spin glass simulations, which revealed the existence of a phase transition in three-dimensional Ising spin glasses.\cite{ogielski} From quite general scaling arguments,\cite{HH} at a continuous phase transition relaxation must be of the form $t^{-x} f(t/\tau(T))$
where $\tau(T)$ diverges as $(T-T_g)^{-z}$, and $f$ is a non-universal function to be determined for each system. For a spin glass the power law exponent $x$ is related to the standard critical exponents through  $x=(d-2+\eta)/2z$ \cite{ogielski}. Here $\eta$ is the Fisher or "anomalous dimension" exponent and $z$ is again the dynamical exponent. The Ising simulations showed that, as T$_{g}$ is approached from above, $q(t)$ is strongly
non-exponential. Ogielski chose to represent $f$ by the stretched exponential or KWW function, familiar in fragile glass dynamics. Excellent fits were obtained with
 $q(t) \propto t^{-x} exp((-t/\tau(T))^{\beta } )$ and $T$ dependent $\tau$ and $\beta$. (The KWW $\beta$ is not to be confused with the critical exponent $\beta$). He and others found a temperature dependent ${\beta}$ tending to near 1/3 at $T_g$ and increasing with $T$. \cite{ogielski,coniglio,ian,goetze} The most important point for our data analysis is that precisely at $T_{g}$ dynamic scaling predicts a pure power law decay for the autocorrelation function : $q(t) \propto  t^{-x}$. This rule is quite general; its functional form does not depend on details such as the Ising character of the spins but the value of $x$ depends on the exponents for the particular system under study. Scaling  therefore gives a unique opportunity for describing the NSE spectra at $T_{g}$ in terms of exponents determined by completely independent dynamic and non-linear macroscopic susceptibility measurements.

       %
       %
\begin{figure}[h]
\resizebox{0.5\textwidth}{!}{%
\includegraphics[scale=0.33]{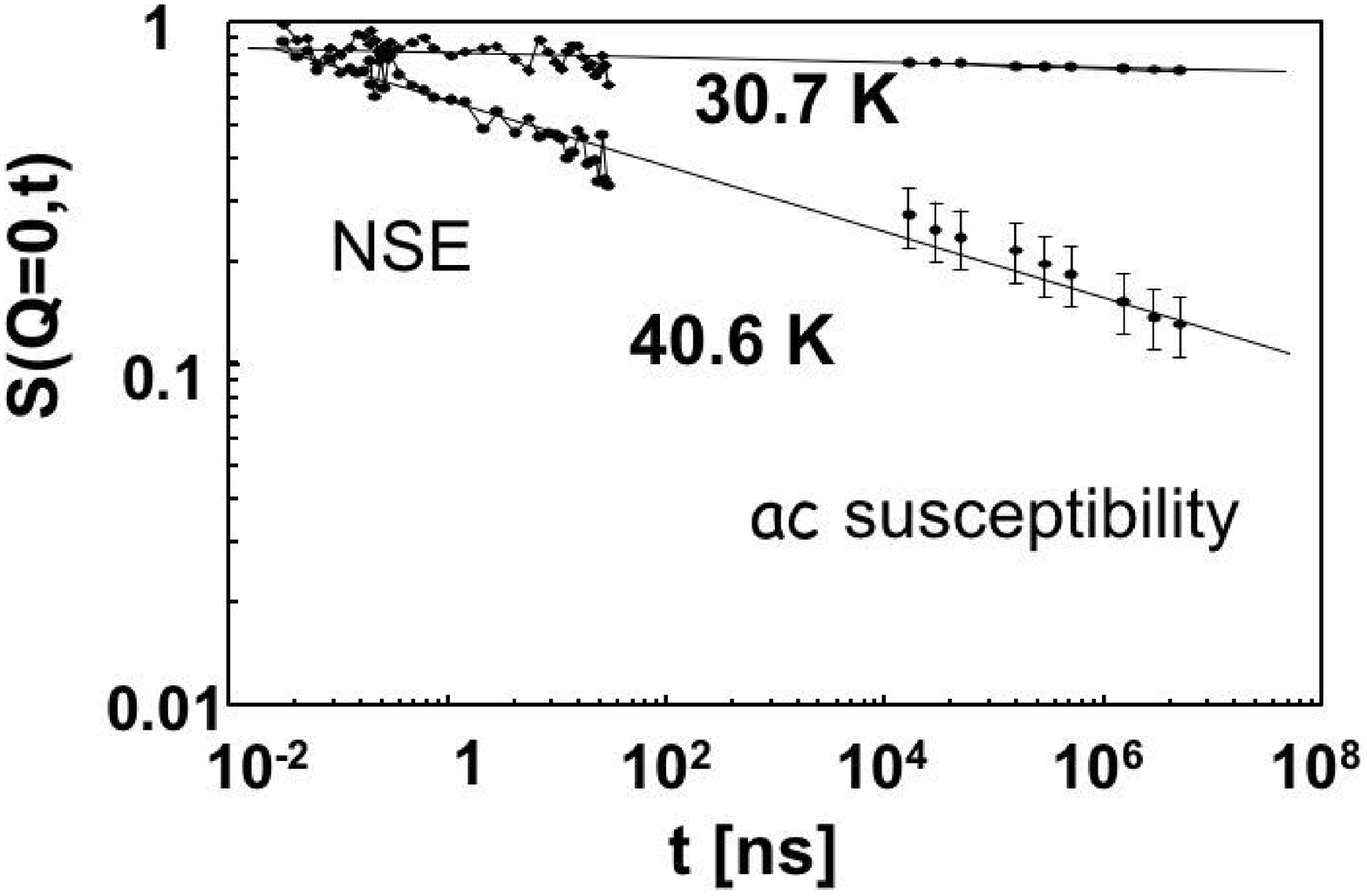}
}
\caption{Combined NSE spectra with s(Q=0,t) values deduced from macroscopic ac susceptibility measurements below T$_{g}$. Also in the close vicinity of T$_{g}$, the pure power law decay of s(Q,t) holds over  an impressively large dynamic range of more  than nine orders of magnitude in time, from the microscopic to the macroscopic times. }
\label{NSE-chi}       
\end{figure}
                               %

As seen on Fig. 1, for T${\leq}$T$_{g}$ we found s(Q,t) $\propto $ t$^{-x} $ and the lines represent the best fits to the data with x=0.116${\pm}$0.007 and 0.025${\pm}$0.005 at T=40.6(${\sim}$T$_{g}$) and 30.7{\nobreakspace}K respectively. Above T$_{g}$ the relaxation is also strongly non-exponential. As shown in Figure 2, pure exponential as well as  stretched  exponential decay (without the power-law prefactor) can be definitively ruled out at all T in the range studied. In fact at 45.6 K, the fit with the Ogielski function leads to $\chi^2$ = 0.28, which is significantly lower than $\chi^2$ = 0.55, the value obtained for the stretched exponential and  $\chi^2$ = 1.67  for the simple exponential respectively. The power law part of the Ogielski function, which holds at short times, describes the main part of the relaxation above T$_{g}$ and the spectra of Figure 1 lead to an accurate determination of the effective power law exponent x as function of T . On the other hand, the stretched exponential only influences the tail of the relaxation so the parameters $\tau(T)$ and ${\beta}$ are obtained here to low accuracy: $\tau(T){\approx}$ 0.3 ns and ${\beta}{\approx}$1 at 55.8 and 50.8 K whereas $\tau(T){\approx}$22 ns and ${\beta}{\approx}$0.66 at 45.7 K.  In  similar systems, muon spin depolarisation measurements which are sensitive to the time range from 10 ns up to about 50 ${\mu}$s (i.e. in the range where the relaxation above T$_{g}$ is mainly described by the stretched exponential) showed that in fact ${\beta}$ approaches about 1/3 at T$_{g}$ \cite{bob,amit} as expected by the simulations. Susceptibility measurements very close to $T_g$ in an Ising spin glass show a similar limiting value of $\beta$.\cite{gunnarsson} Furthermore, recent magnetic-field dependent muon spin depolarisation measurements on several AgMn spin glasses, analysed assuming an Ogielski-like decay of the correlations, lead to values of $x$ tending to ${\sim}$0.15 at T$_{g}$ in agreement with our results.\cite{amit}

The large dynamic range covered by our data and their accuracy allow us to distinguish between a simple stretched exponential decay, the Ogielski function with $\beta = 1$ (power law times simple exponential), and the full Ogielski function; the full function is needed to give an acceptable fit of the data over the whole time and temperature range.  

%
\begin{figure}[t]
\resizebox{0.5\textwidth}{!}{%
\includegraphics[scale=0.35]{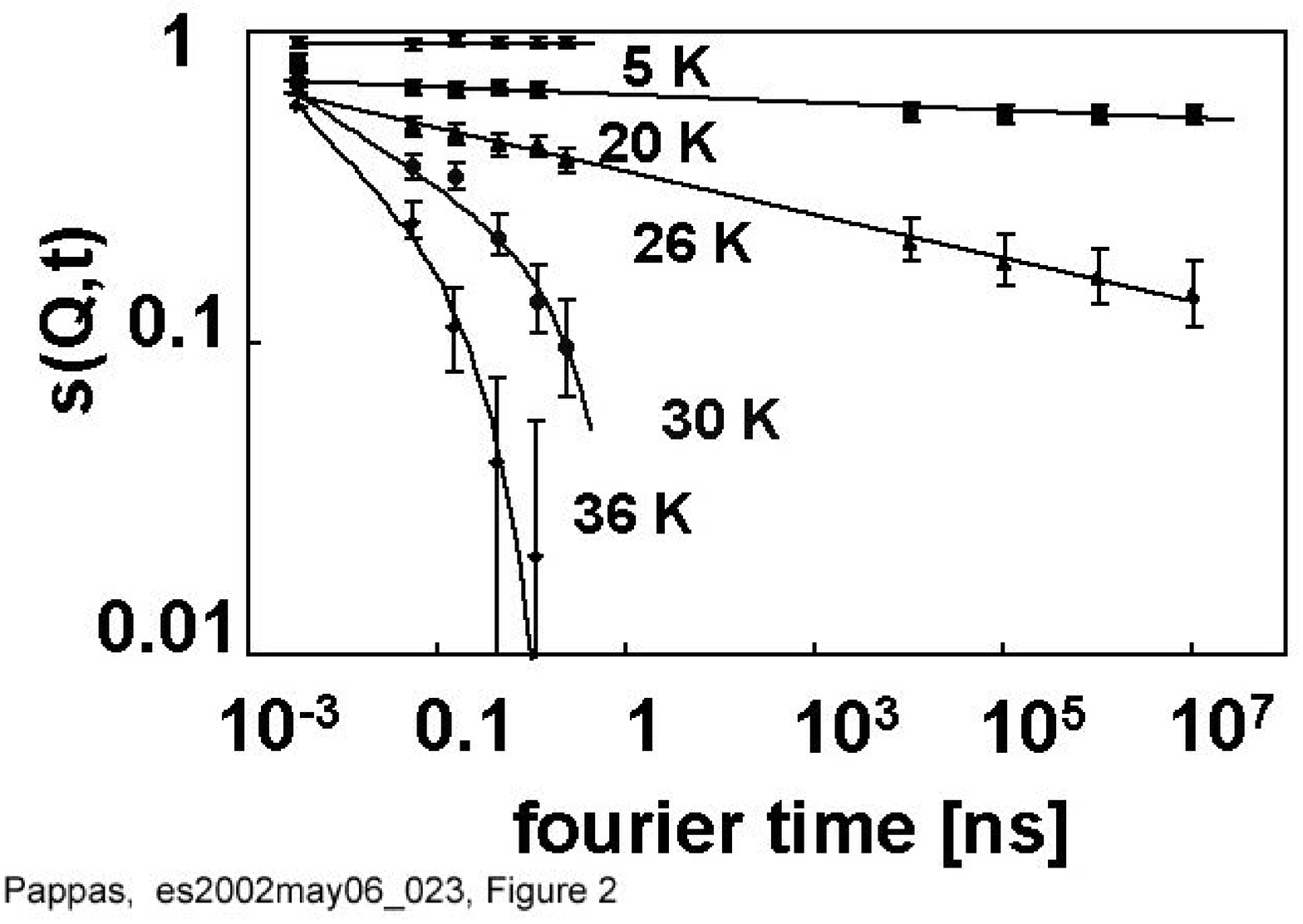}
}
\caption{Revisited analysis of the historical  s(Q,t) data on CuMn 5\% (5). Below T$_{g}$ (${\sim}$27.5 K) the NSE spectra
were combined with values calculated from the dynamic susceptibility (10$^{4} {\geq}$ t ${\geq}$ 10$^{7}${\nobreakspace}ns).
The data are plotted in a log-log scale and the continuous lines correspond to a simple
power law decay below T$_{g}$ and to the Ogielski function above T$_{g}$.}
\label{cumn}       
\end{figure}
                               %

%
\begin{figure}[hb]
\resizebox{0.5\textwidth}{!}{%
\includegraphics[scale=0.35]{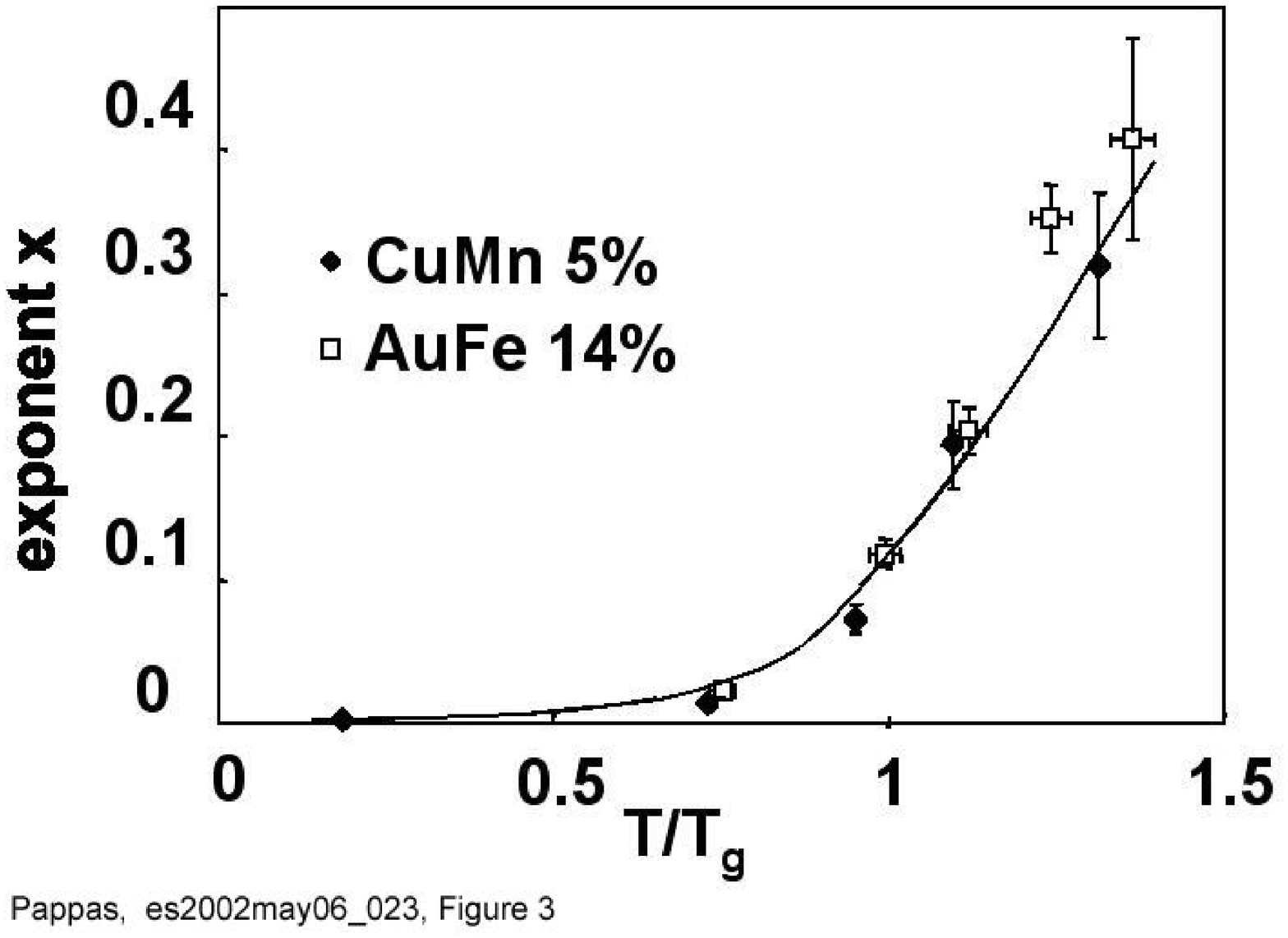}
}
\caption{Temperature dependence of the exponent x obtained
by fitting the NSE spectra to a simple power low decay below
T$_{g}$ and to the Ogielski function above T$_{g}$. At T$_{g}$
x${\sim}$0.12 as predicted by dynamic scaling.  }
\label{exponentx}       
\end{figure}
                               %

In spin glasses the normalized $s(Q,t)$ does not vary with $Q$, in dramatic contrast to the strong Q-dependence of the dynamics in ferromagnets. This remarkable Q-independence of the relaxation in spin glasses \cite{feri-murani}  implies that the NSE s(Q,t) can be identified with q(t) and analysed in the frame of dynamic scaling.  The NSE spectra of Au$_{0.86}$Fe$_{0.14}$ collected on IN15 for $Q=0.4$ and $0.8${\nobreakspace}nm$^{-1}$ and on SPAN for 0.6{\nobreakspace}nm$^{-1}{\leq}$Q${\leq}$2.6{\nobreakspace}nm$^{-1}$ confirmed this behaviour. Fig. 3 compares,  in a lin-log plot, spectra collected at 45.6 and 40.6 K for   
Q= 0.04{\nobreakspace}\AA$^{-1}$ and 0.08{\nobreakspace}\AA$^{-1}$ respectively.  
The data almost  overlap although the Q-values differ by a factor of two and the magnetic 
 intensity decreases by almost a factor of 3, which
 explains the larger error bars of the data set  at 0.08{\nobreakspace}\AA$^{-1}$. 
The Q-independence of s(Q,t) was an important check of the high quality of our samples in a concentration range close to ferromagnetism. It also implies that s(Q,t) can be directly related to the macroscopic ac susceptibility  $\chi$(Q=0, $\omega$) = S(Q=0)[1-s(Q=0,t)]/kT.\cite{feri} The meaning of this equation is quite simple: $s(Q,t)$ is the fraction of the total magnetic response S(Q) which does not relax before the time t, in other words $s(Q,t)$ is the part of S(Q) which cannot respond to a driving field of frequency $1/t$. Figure 4 shows the NSE spectra for $T\leq T_g$ combined with $s(Q,t)$ values deduced from macroscopic dynamic (a.c.) susceptibility measurements on the same sample. The data follow the power law decay over an impressively large range of at least 9 orders of magnitude in time.

The impressive similarity between the experimental s(Q,t) and the decay of q(t) found in large scale simulations and more particularly the simple power law decay found at and below T$_{g}$ incited us to revisit the historical CuMn 5\% data. Below T$_{g}$, the spectra of CuMn 5\% were combined with macroscopic dynamic (ac) susceptibility measurements and thus covered more than 9 orders of magnitude in time down to the microscopic time scale.\cite{feri} These data, plotted now on a log-log scale, also reveal a power law decay below T$_{g}$ over an impressively large range of at least 9 orders of magnitude in time as shown in Fig. 5. Above T$_{g}$, which is at 27.5{\nobreakspace}K, the power law holds only at short times and the decay can be described by the Ogielski function with ${\beta} {\approx}$1 just as in AuFe 14\%.

The values of the exponent x, plotted versus the reduced temperature T/T$_{g}$ in Fig. 6, are similar for both metallic systems. This confirms the similarity of the dynamic behaviour of AuFe, AgMn and CuMn systems around T$_{g}$, pointed out by Uemura after comparing the NSE spectra of CuMn 5\% with ${\mu}$SR spectra of AgMn and AuFe.\cite{uemura} As already mentioned, dynamic scaling relates the value of x at T$_{g}$ to the static and dynamic critical exponents, which can be determined completely independently from non-linear and ac macroscopic susceptibility measurements. Well established values of these exponents are available for a AgMn spin glass \cite{levy} and for CuMn and AgMn spin glasses doped with Au,\cite{nadine} in agreement with more recent data on AuFe 14\%. The static exponents lead to ${\eta}$= 2-${\gamma}/{\nu}$ = 0.23${\pm}$0.3 and the dynamic exponent z=5.3${\pm}$0.8 was determined by low frequency susceptibility measurements. From these values we calculate x=0.116${\pm}$0.026 in excellent agreement with x=0.116${\pm}$0.007,  the value we have observed on Au$_{0.86}$Fe$_{0.14}$ at T$_{g}$. It is important to note that the value of $x$ in these basically Heisenberg spin glasses is considerably larger than that seen numerically \cite{ogielski} or experimentally \cite{gunnarsson} in Ising spin glasses where $x {\approx} 0.07$.

Our results represent the first verification of scaling in spin glasses relating quantities of very different nature and measured by different methods on very different time scales, namely the microscopic time dependence of the autocorrelation function at times between the microscopic time and $10^{-7}$ seconds on the one hand, and the macroscopic a.c. susceptibility at time scales greater than $10^{-3}$ seconds together with the static non-linear susceptibility on the other. This agreement constitutes a most compelling evidence for a true phase transition in spin glasses at T$_{g}$. In addition, this demonstration of critical relaxation characterisic of a phase transition also implies that the entire non-exponential temporal relaxation in these spin glasses is an intrinsic, homogeneous feature, as already evidenced by Uemura after comparing ${\mu}$sR, NSE and dynamic susceptibility data.\cite{uemura}  This conclusion is in obvious contrast to the clear evidence for the heterogeneous origin of the non-exponential relaxation in some structural glasses, where also no clear signature of a phase transition could be found.\cite{glasses} The clear evidence for a phase transition with a non-conventional order parameter and intrinsic, homogeneous
non-exponential relaxation in spin glasses, a member of the large family of glassy systems, which also include structural glasses,
other non-crystalline systems and living matter, is of particular importance in view of the unsolved central question of the nature of the
glass transition in general.
\\
\\

\textbf{ACKNOWLEDGMENTS}
\\

C. Pappas would like to acknowledge fruitful discussions with  R. Cywinski and S. Kilcoyne, who also made possible the susceptibility measurements at the Physics \& Astronomy Department of the University of Leeds.  P Manuel would like to acknowledge postdoctoral support from EPSRC Research Grant GR/M77260.

\pagebreak

\end{document}